\def\sv{\vskip 0.05in}
\def\s{\vskip 0.15in}

\documentclass[prb,twocolumn]{revtex4}
\newcommand{\formula}{${\rm LiCu_{2}O_{2}}\;$}
\usepackage{graphicx}% Include figure files
\usepackage{dcolumn}% Align table columns on decimal point
\usepackage{bm}% bold math
\usepackage{amsmath}
\usepackage{multirow}
\begin{document}

\title{Investigation of the Structural and Dynamical Properties of the (001) Surface of \textup{LiCu$_{2}$O$_{2}$}}

\author{Yangyang Yao$^1$, Xuetao Zhu$^1$,H. C. Hsu$^2$, F. C. Chou$^2$ and M. El-Batanouny$^1$}\sv

\affiliation{\sl{$^1$ Physics Department, Boston University, Boston,
MA 02215\\$^2$ Center of Condensed Matter Sciences, National Taiwan
University, Taipei 10617, Taiwan}}
\begin{abstract}
We report on studies of the structure and dynamics of the (001)
surface of single crystal \formula, investigated by He beam
scattering at room temperature, and with lattice dynamical models.
The best fit surface corrugation to measured diffraction patterns
shows that the surface termination is exclusively a $\rm
Li^{+1}Cu^{+2}O^{-2}$ plane. Lattice dynamics fits to inelastic He
scattering spectra reveal the presence of two low-lying surface
phonon modes, identified with the motion of $\rm{Cu^{+2}}$,
$\rm{Li^{+1}}$ surface ions normal to the surface.
\end{abstract}

 \maketitle
\section{Introduction}
\formula continues to attract considerable attention because of the
unique physical properties it exhibits. Initially, interest in this
system was stimulated by the presence of double-chain ladders of
Cu$^{+2}$O, which presented a prototype quasi-one-dimensional (QOD)
spin-1/2 quantum magnetic system with competing magnetic
interactions. Competing magnetic interactions in the double-chain
ladder were known to give rise to geometric frustration, which in
turn is manifest in an ordered incommensurate helimagnetic phase at
low temperatures. It was expected that the presence of S=1/2 spins
would give rise to strongly competing commensurate quantum spin
fluctuations that tend to suppress the transition temperature
\cite{bi8,bi15,bi12}. More recently, it was discovered that this
system exhibits ferroelectricity upon the emergence of the spiral
magnetic order. This renders \formula as the second cuprate to join
the list of multiferroics \cite{bi28,bi29,bi30,bi16}.

A series of successive magnetic phases have been reported at low
temperatures \cite{bi8,bi10,bi17}: Electron spin resonance (ESR)
measurements revealed the presence of a dimerized spin-singlet state
at T$>23$K, with an energy gap of $\Delta$=72K between the
spin-singlet ground state and the first spin-triplet excited state.
Magnetic neutron scattering measurements confirmed the existence of
a QOD spin-ordered helical phase in the temperature range $9$K
$<$T$<23$K\cite{bi10,bi15}, and a collinear anti-ferromagnetic (AFM)
phase was suggested to exist below $9$K\cite{bi7,bi17,bi14}. The
presence of the QOD magnetic ordering was attributed to the presence
of impurities in the chains: The observation of the classical
helical phase in LiCu$_2$O$_2$ was ascribed to substitutional Li$^+$
defects that tend to suppress effective long-range
one-dimensionality \cite{bi8}.

\begin{figure}[h!]
  \includegraphics[width=2.75in]{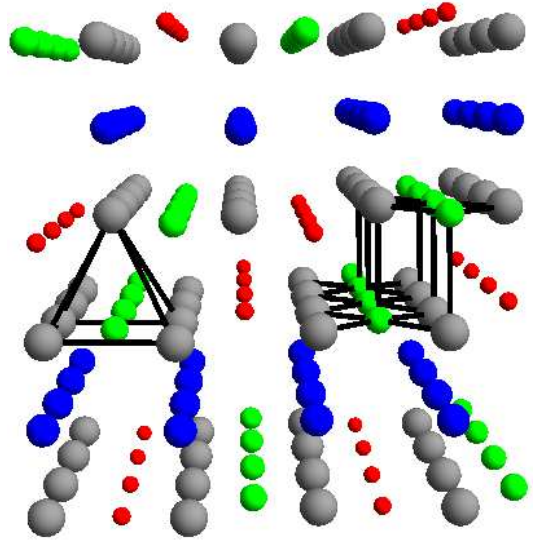}\\
\caption{ Crystal structure of LiCu$_2$O$_2$. The color code is: Li
(red), O (gray), Cu$^{+1}$ (blue), Cu$^{+2}$
(green)}\label{Li1}\end{figure}\sv

\formula is a quasi-1D insulator. It has a layered charge-ordered
orthorhombic crystal structure belonging to the {\sl Pnma} space
group; the primitive cell has lattice constants $a$=5.73$\rm \AA$,
$b$=2.86$\rm \AA$, $c$=12.47$\rm\AA$ respectively\cite{bi5}. It is a
mixed-valent compound with copper ions in the Cu$^{+2}$ and
Cu$^{+1}$ valance states. The magnetic Cu$^{+2}$ (S=1/2) ions are
located at the center of edge-sharing CuO$_4$ plaquettes which form
infinite chains along the crystallographic $b$-axis. Coplanar chains
are connected by chains of Li ions along the $a$-axis, and two such
planes form double-layers parallel to the $ab$-plane, as shown in
figure \ref{Li1}. The QOD spin arrangement is due to these double
chains of Cu$^{+2}$ ions that run along the crystallographic $b$
axis. The period of each "leg" of the double spin chains is equal to
$b$. The two legs are offset by b/2 relative to each other. Along
the c-axis, each double-layer is separated from its double-layer
neighbors by magnetically inert Cu$^{+1}$ planes.

Despite the extensive studies of bulk structural and magnetic
properties of \formula crystals cited above, no investigation of its
surface properties has been reported in the literature. It is known
that \formula crystals easily peel along the (001) surface. However,
since these crystals consist of alternating double-layers of
Li$^{+1}$Cu$^{+2}$O$^{-2}_2$ and single layers of Cu$^{+1}$, it
remains to be determined whether the cleavage would result in the
coexistence of Li$^{+1}$Cu$^{+2}$O$^{-2}_2$ and Cu$^{+1}$ surfaces,
or in an exclusive presence of one of these layer types. If the
latter scenario occurs, then it would dictate that the double-layer
must split in order to provide complete coverage of the two newly
exposed surfaces. The double-layer splitting means the splitting of
the double-chain and the presence of a single-chain, which is
expected to display different magnetic behavior from that observed
in the double-chain.

In this paper we present the results of extensive studies of the
(001) surface of \formula, using the experimental techniques of
elastic and inelastic helium scattering aimed at determining its
surface structural and dynamical properties, respectively.
Furthermore, we used lattice dynamical models with slab geometries
to interpret the surface phonon dispersion curves derived from the
measured inelastic scattering spectra. The surface corrugation
topography, derived from a large set of diffraction patterns,
clearly shows that the surface termination is exclusively
Li$^{+1}$Cu$^{+2}$O$^{-2}_2$. Moreover, empirical lattice dynamics
models, with slab geometries based on such termination and fit to
the inelastic experimental results, reveal two low-lying dispersion
curves with polarizations normal to the surface, one involves the
Cu$^{+2}$ and the second involves Li ions. In section II the
experimental setup and procedures are discussed, and the results and
conclusion are presented in section III.

\section{Experimental Setup and Procedure}
Single crystals of \formula with high Li content of $\simeq 0.99\pm
0.03$ were grown by the floating-zone method. The Li content was
determined accurately through combined iodometric titration and
thermogravimetric methods. This ruled out the possibility of
chemical disorder between Li and Cu ions. Details of the growth
procedures and stoichiometry confermation are given in Ref.
\cite{bi31}. Typical crystal samples used were about
3mm$\times$3mm$\times$2mm in size, with its exposed surface parallel
to the $ab$-plane. The crystals were attached to an OFHC copper
sample-holder by conductive silver epoxy. A cleaving (peeling) post
was attached to the top sample surface in the similar way. The
prepared sample holder was mounted on a sample manipulator equipped
with XYZ motions as well as polar and azimuthal rotations. The
pressure in the Ultra-High Vacuum (UHV) chamber was maintained at
$10^{-10}$ torr throughout the experiment to ensure the cleanliness
of the sample surface during measurement performance. In situ
cleaving under UHV conditions was effected by knocking off the
cleaving post. Immediately after cleaving, the quality of the
long-range ordering on the surface was confirmed by the appearance
of sharp diffraction LEED spots.

A supersonic mono-energetic collimated helium beam, with velocity
resolution better than 1.4 $\%$, was generated by a nozzle-skimmer
assembly and 2mm diameter collimating slits. The average beam
velocity was varied by attaching the nozzle reservoir to a
closed-cycle helium refrigerator, and controlling the reservoir
temperature with the aid of a digital temperature controller
(Scientific Instruments Model 9700) and a diode sensor attached to
the reservoir. As a result, the beam energy can be varied in the
range 65meV to 21meV by varying the nozzle temperature from 300K to
110K. Polar rotation of the sample was used to vary the incident
angle $\theta_{i}$ with respect to the surface normal, while the
azimuthal rotation was employed to align the scattering plane along
a high-symmetry surface crystallographic direction. The scattered He
beam was collected by an angle-resolved detector mounted on a
two-axis goniometer, which allows the scattered angle $\theta_{f}$
to be varied independently from $\theta_{i}$ \cite{bi18}, and allows
in- and out-of the scattering-plane measurements. The
detector\cite{bi26} is comprised of an electron gun and a
multichannel plate (MCP) electron multiplier. The electron gun
generates a well-collimated, monoenergetic electron beam crossing
the He beam at right angles. The energy of the electron beam is
tuned to excite the He atoms to their first excited metastable state
(2 $^3$S He*) upon impact. Deexcitation of a He* atom at the surface
of the MCP leads to the ejection of an electron which generates an
electron cascade that is then collected by the anode of the
multiplier. By electronically pulsing the electron gun, a gate
function is created for time-of-flight (TOF) measurements in the
inelastic HAS mode. The details of the detection scheme are given in
Ref 14.
%\cite{bi26}.
 All measurements were performed with the sample surface at room temperature.

 By writing the He-atom wave vector as ${\bf k}=({\bf K},k_z)$, where ${\bf K}$ is
 the component parallel to the surface, conservation of momentum and
energy for in-the-scattering-plane geometry can be expressed as
\begin{align}\label{1a}
\Delta{\bf K}\,&=\,{\bf G}+{\bf Q}\,=\,k_{f}\sin\theta_{f}-k_{i}\sin\theta_{i}\\\label{2a}
\Delta E\,&=\,\hbar\omega({\bf Q})\,=\,E_{i}\,\left[\biggl(\frac{\sin\theta_{i}+\Delta K/k_{i}}{\sin\theta_{f}}\biggr)^{2}-1\right]
\end{align}
where subscripts i and f denote incident and scattered beams,
respectively, and $\Delta{\bf K}$ is the momentum transfer parallel
to the surface. ${\bf G}$ is a surface reciprocal-lattice vector,
${\bf Q}$ is the surface phonon wave vector, and $\hbar\omega({\bf
Q})$  is the corresponding surface phonon energy.
$E_i=\hbar^2k_i^2/2M$, where $M$ is the mass of a He atom. By
eliminating ${\bf k}_f$ from the above equations, one obtains the
so-called scan curve relations which are the locus of all the
allowed $\Delta{\bf K}$ and $\Delta E$ as dictated by the
conservation relations,
\begin{equation} \Delta E\,=\,E_i\,\left[\left(\frac{\sin\theta_i + \Delta{\bf K}/k_i}{\sin\theta_f}\right)^2-1\right].
\end{equation}
The intersections of these scan curves with the phonon dispersion
curves define the kinematically allowed inelastic events for a fixed
geometric arrangement. Thus, by systematically changing
$E_i,\,\theta_i$, and $\theta_f$, the entire dispersion curves can
be constructed.

\section{{\bf Results and Discussion}}
\subsection{Elastic He Scattering And Surface Structure}
\begin{figure}
  \includegraphics[width=3.5in]{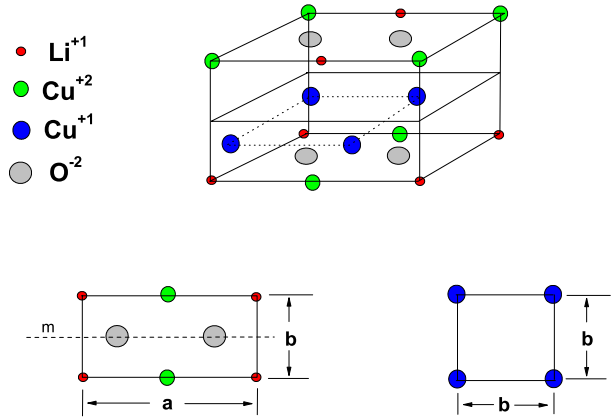}\\
  \caption{Top: Half of a unit cell of \formula. Bottom: Two candidates
 for surface termination: $\rm{Cu^{+1}}$ layer with a square lattice
(bottom right), or a Li$^{+1}$Cu$^{+2}$O$^{-2}_2$ layer (bottom
left). }\label{fg2a}
\end{figure}

Diffraction patterns were collected from several crystal samples at
a temperature of 300 K for several scattering conditions:
\begin{enumerate}
\item incident He wave numbers, $k_{i}$, in the range 6.42$\rm{\AA}^{-1}$ and 11.12$\rm{\AA}^{-1}$,
\item incident angles, $\theta_{i}$, between $30^{\circ}$ and $50^{\circ}$, and
\item two high-symmetry azimuthal surface orientations, $a$ and $b$, separated by $90^{\circ}$.\end{enumerate}
As was mentioned above, there are two candidates for the surface
termination, which are shown in Figure.\ref{fg2a}, the
Li$^{+1}$Cu$^{+2}$O$^{-2}_2$ layer and the Cu$^{+1}$ layer. The
latter has a square lattice with lattice constant $b$, and $C_{\rm
4v}$ symmetry. Thus, the diffraction peaks that correspond to odd
multiples of $2\pi/a$ should be absent for pure Cu$^{+1}$ surface
termination, which contradicts the fact that they do appear in all
the measured diffraction patterns. Alternatively, for
Li$^{+1}$Cu$^{+2}$O$^{-2}_2$ termination, the well known crystal
twinning\cite{bi16,bi30,bi8} would imply that the two azimuthal
orientations would give the same diffraction pattern. This means
that we obtain a superposition of diffraction patterns from
\textbf{a}$\rm{\langle10\rangle}$ and
\textbf{b}$\rm{\langle01\rangle}$ directions. Moreover, since for
\formula $a\,\simeq\,2b$, the diffraction peaks associated with the
$b$-direction will lie very close to the even-order peaks obtained
for the $a$-direction.

\begin{figure}[h]
  \includegraphics[width=3.5in]{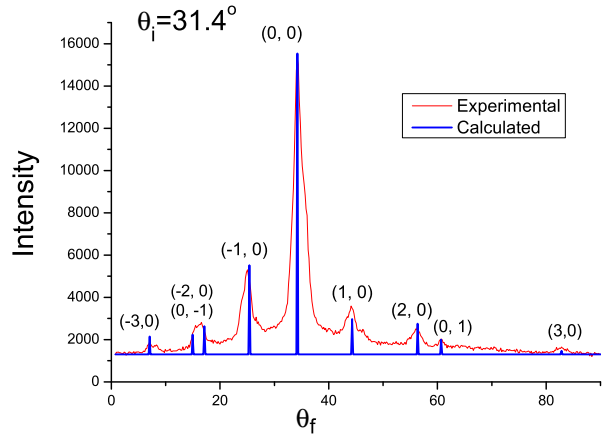}\\
  \caption{The experimental diffraction pattern with $k_{i}=7.28 \rm{\AA}^{-1}$, $\theta_{i}=31.4^{\circ}$ at T=300K (red), and the calculated diffraction intensities (blue vertical bars) }\label{fg1}
\end{figure}

Figure \ref{fg1} shows a typical diffraction pattern recorded for
$k_{i}=7.28 \rm{\AA}^{-1}$ and $\theta_{i}=31.4^{\circ}$.  Although
the twinning makes the situation a little complicated, luckily, the
angular resolution in the diffraction pattern allows us to clearly
separate the peak positions along the two directions, as indicated
by peaks (2,0) and (0,1) in figure \ref{fg1}. The coexistence of
both surface terminations would imply that the intensity of the
(1,0) and (0,1) peaks should be comparable, but, in fact, the (0,1)
peak has exhibited appreciably lower intensity than the (1,0) peak.
Hence, we shall follow the scenario of a purely
Li$^{+1}$Cu$^{+2}$O$^{-2}_2$ termination.

The surface periodicity along \textbf{a} and \textbf{b} is derived
from the positions of their corresponding diffraction peaks using
the relation
\begin{equation}
\left|{\bf{k}_{i}}\right|\,(\sin{\theta_{i}}-\sin{\theta_{\bf G}})\,=\,\left|{\bf G}\right|
\end{equation}
where $\bf{k}_{i}$ is the incident beam wave-vector, ${\bf G}$ a
surface reciprocal lattice vector, $\theta_{\bf G}$ the
corresponding scattering angle. Analysis of all the collected
diffraction patterns yield the surface lattice constant values:
$a_{s}=5.76 \pm 0.05\rm \AA$ and $b_{s}=2.91 \pm 0.08\rm{\AA}$,
which compares quite well with the bulk values of a=5.73$\rm \AA$,
b=2.86$\rm \AA$ and, thus, precludes the presence of diffraction
satellites and surface reconstruction.

In order to determine the topology of the surface primitive cell we
used the \textit{hard corrugated wall model} within the
\textit{eikonal approximation} to calculate the elastic scattering
intensities\cite{bi24}. Here, the surface topology is defined in
terms of a surface corrugation function
$\zeta(\mathbf{R})$\cite{bi25}, where ${\bf R}$ denotes a surface
position vector and $\zeta$ the corrugation height at that position.
In this framework, the scattering amplitude of the diffraction peak
${\bf G}$ is given by\cite{bi25}
\begin{equation}\label{2}
A_\mathbf{G}=-\frac{1}{S}\int_{\rm{u.c.}}\;e^{i[\mathbf{G}\cdot\mathbf{R}+q_{\mathbf{G}z}\zeta(\mathbf{R})]}\,d\mathbf{R}
\end{equation}
where $q_{\mathbf{G}z}=k_{iz}-k_{\mathbf{G}z}$, with $k_{\mathbf{G}z}=k_i\,\cos\theta_{\bf G}$. The integration is carried over the unit cell (u.c.) of area $S$. Since $\zeta(\mathbf{R})$ should have the surface periodicity, we write
\begin{equation}\label{3}
\zeta(\mathbf{R})=\sum_{\mathbf{G}}\;\zeta_\mathbf{G}\,\textup{exp}\,(i\mathbf{G}\cdot\mathbf{R})
\end{equation} We make use of the reflection symmetry perpendicular to the $b$-direction ($y$-axis) of the $\textup{Li}\textup{Cu}_2\textup{O}_2$ surface to simplify equation (\ref{3}), and write
\begin{align}
\zeta(x,y)&=\sum_{n_1,n_2}\;\zeta_{n_1,n_2}\,\sin(n_1b_1x)\cos(n_2b_2y)\nonumber\\
&\hspace{1pt}+\sum_{n_3,n_4}\,\zeta_{n_3,n_4}\,\cos(n_3b_1x)\cos(n_4b_2y)
\end{align}
where $b_1=\frac{2\pi}{a}$, $b_2=\frac{2\pi}{b}$, are the surface reciprocal lattice vector basis.

The goal now is to determine the coefficients $\zeta_\mathbf{G}$.
The following iterative fitting scheme was adopted\cite{bi27}.
Initially, the magnitudes $\left|A_\mathbf{G}\right|$ are determined
from the experimental diffraction pattern using the geometric
relation
\begin{equation}
P_\mathbf{G}=\frac{|k_{\mathbf{G}z}|}{|k_{iz}|}\left|A_\mathbf{G}\right|^2
\end{equation}  The experimental intensities are normalized to satisfy the unitarity condition
\begin{equation}
\sum_{\mathbf{G}}P_\mathbf{G}=1
\end{equation}
Next, the eikonal equation\cite{bi24,bi25,bi27}
\begin{equation}\label{6}
\sum_{\bf G}\;A_\mathbf{G}\,e^{i{\bf G}\cdot{\bf R}}\,e^{ik_{{\bf G}_z}\,\zeta({\bf R})}\,=\,-e^{ik_{iz}\zeta({\bf R})}
\end{equation} where $A_\mathbf{G}=\left|A_\mathbf{G}\right|\,e^{i\phi_{\bf G}}$, is
used to determine $\zeta({\bf R})$. $\phi_{\bf G}$ is a diffraction
phase angle to be determined. In the first iteration we set $A_\mathbf{G}=\left|A_\mathbf{G}\right|$, which yields
$$\zeta_0({\bf R})\,=\,\frac{\ln\left(-\sum_{\bf G}\;A_\mathbf{G}\,e^{i{\bf G}\cdot{\bf R}}\right)}{2ik_{iz}},$$
where we replaced $k_{{\bf G}_z}$ by $-k_{iz}$ in equation (\ref{6}). Further iterations involved varying the amplitudes $\zeta_{\bf G}$ around the values obtained from $\zeta_0({\bf  R})$.

\begin{figure}[h]
  \includegraphics[width=3.5in]{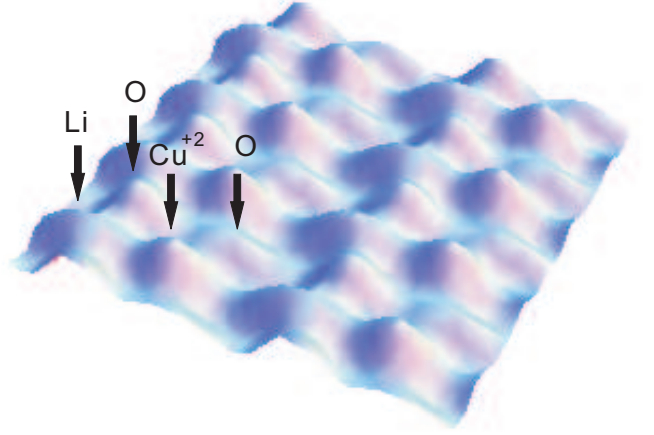}\\
  \caption{Calculated surface corrugation function}\label{fg2}
\end{figure}

 After each iteration the reliability factor $R$
\begin{equation}
R=\frac{1}{N}\sqrt{\sum_\mathbf{G}(P_\mathbf{G}^{\textup{exp}}-P_\mathbf{G}^{\textup{calc}})^2}
\end{equation}
where N is the number of diffraction peaks, was calculated. The
iterations were terminated when an acceptable small value of $R$ was
reached. The corrugation function with best-fit parameters,
$R=0.003$, is
\begin{align}
\zeta(x,y)&=-0.06\cos(b_1x)-0.05\cos(2b_1x)+0.02\cos(3b_1x)\nonumber\\
&\hspace{1pt}+0.03\sin(b_1x)-0.03\cos(b_2y)+0.05\cos(2b_2y)\nonumber\\
&\hspace{1pt}-0.07\cos(b_2y)\sin(2b_1x)
\end{align}
It is plotted in figure \ref{fg2}, and the corresponding calculated
diffraction peak intensities are shown as blue vertical bars in
figure \ref{fg1}. The locations of the Li, Cu$^{+2}$,and the two O
ions have been identified by comparing the positions of the
corrugation maxima to ionic positions in the unit cell; they are
indicated in the figure.
\subsection{Inelastic Measurements And Shell-Model Calculations}
\begin{figure}
\includegraphics[width=2.5in]{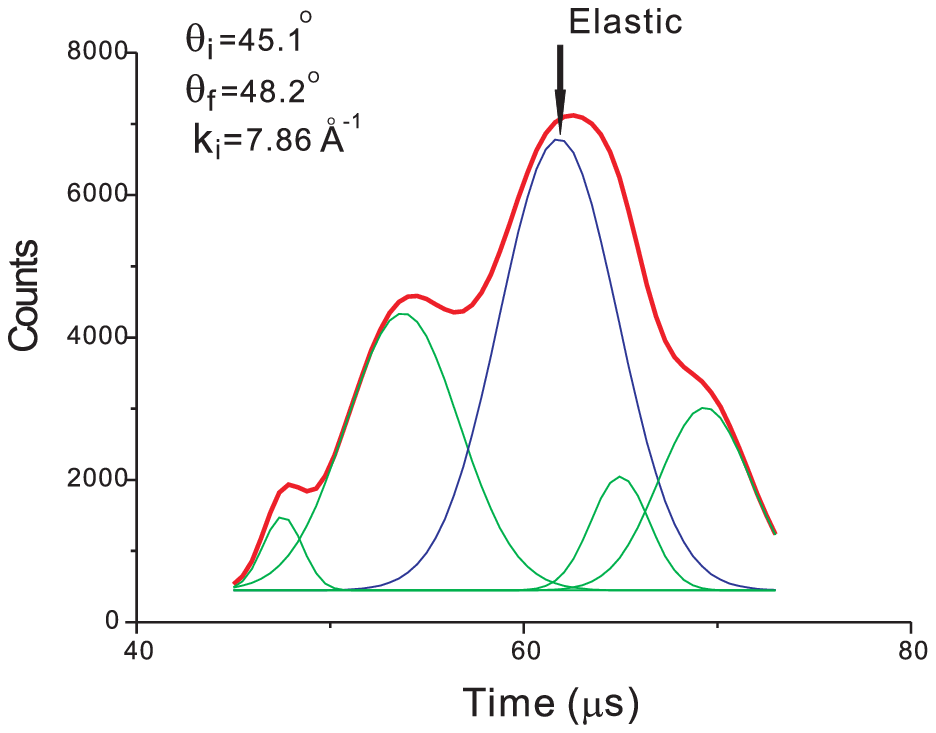}\\\sv  \includegraphics[width=2.5in]{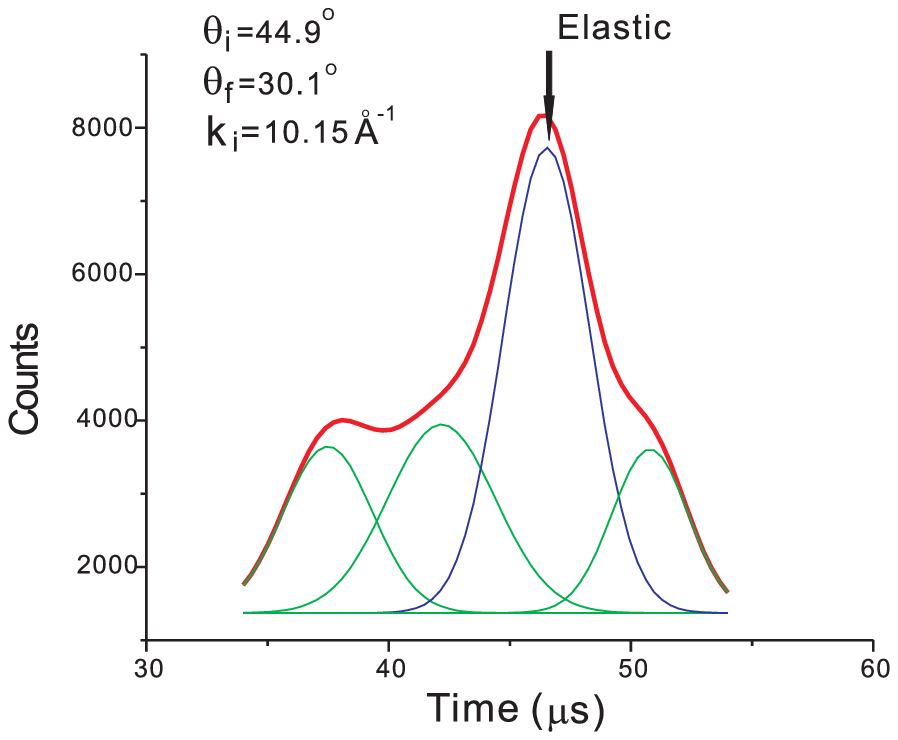}\\\sv
\caption{Typical time-of-flight spectra. The experimental curve (red) is resolved into inelastic peaks (green) and the diffuse elastic peak (blue), which is also indicated by arrows.}\label{fg3}
\end{figure}
Inelastic scattering measurements were carried out for
in-the-scattering-plane geometry. Two typical TOF spectra showing
diffusive elastic as well as inelastic peaks are shown in figure
\ref{fg3}. The data were collected using beam energies in the range
of 25-65 meV. The high-energy He beam, $E_i=65$ meV, was used to
probe the high-frequency surface phonon modes. The energy resolution
for the lowest-energy beam was about 3 meV. Phonon energies and
momenta were calculated from the TOF peak positions, after a
Gaussian fit to the peaks, using Eqs. (\ref{1a}) and (\ref{2a}). To
characterize the ensuing phonon dispersion points, lattice-dynamical
analysis for the bulk and surface (slab calculation) was carried
out.
\subsection*{1. Static equilibrium and bulk lattice dynamics}
In lattice-dynamical studies of complex systems such as \formula,
the construction of a dynamical matrix based on pair potentials
rather than mere force constants is far more advantageous in the
sense that the underlying physics is transparent and many of the
potential parameters for similar pair interactions can be
transferred from one compound to another. This is especially useful
when experimental phonon frequencies are limited to a few bulk
infrared (IR) and Raman active phonon modes at the center of the
Brillouin zone (BZ)\cite{bi23,bi18}, and no neutron-scattering data
for phonon dispersion is available. Bulk lattice dynamics
calculations, based on the shell-model, was used to obtain an
empirical fit to existing IR and Raman data.
\begin{table}[h!]\caption{\label{pot}Best-fit Born-Meyer and shell parameters.}\sv
   \begin{tabular}{c|c|c|c}\hline\hline\multicolumn{3}{c}{}\\\multicolumn{4}{c}{Born-Meyer Potential Parameters} \\[3pt]\hline&&&\\[3pt]Bond& \small a(eV)& b(${\rm \AA^{-1}}$) & \small c(eV${\rm\AA^{6}}$)\\[3pt] \hline&&&\\
  Cu$^{+1}$-O &18705.6&2.9&$-$\\[3pt]
  Cu$^{+2}$-O&10000&4.4&$-$\\[3pt]
  Cu$^{+1}$-Cu$^{+2}$&3490.95&3.0&$-$ \\[3pt]
  Cu$^{+1}$-Cu$^{+1}$&18853 &3.0&$-$\\[3pt]
  Cu$^{+2}$-Cu$^{+2}$&6598.44&3.0&$-$ \\[3pt]
  ${\rm O-O}$& 2146.30 &3.4 &50 \\[3pt]
  ${\rm Li-O^{a}}$ &840 &3.0&$-$\\[3pt] \hline\end{tabular}\sv
\begin{tabular}{c|c|c|c}\hline\hline\multicolumn{3}{c}{}\\\multicolumn{4}{c}{shell model parameters} \\[3pt]\hline&&&\\
  Ion  &Z(e)  & Y(e) &K(eV/${\rm \AA^{2}}$) \\[3pt] \hline&&&\\
  Li  & +1   & 1.5  & 5  \\[3pt]
  Cu$^{+1}$ & +1   & 1.7  &20  \\[3pt]
  Cu$^{+2}$ & +2   & 2.9  &30  \\[3pt]
  O$^{b}$ &-2    &-3.1  &40  \\[3pt] \hline\end{tabular}\s
$^{a}$ref.\cite{bi23},\,$^{b}$ ref.\cite{bi18}
\end{table}
The model incorporates two-body central potentials, namely the
Coulomb potential $V^C_{ij}(r)=Z_iZ_je^2/r$ for the long-range
interactions and either Born-Mayer $V_{ij}(r)=a_{ij}e^{-b_{ij}r}$ or
Buckingham  $V_{ij}(r)=a_{ij}\exp(-b_{ij}r)-c_{ij}/r^{6}$ potentials
for the short-range interactions. The ionic shells are characterized
by the ionic charge Z, the shell charge Y, and the intra-ion
shell-core force constant K. Some of the initial short-range pair
potentials were obtained from the literature:
(Cu$^{+2}$-O)\cite{bi18}, (O-O)\cite{bi18} and (Li-O)\cite{bi23}.
The remaining pair potential parameters were set to satisfy the
static equilibrium conditions\cite{bi11} which state that the forces
on the particles in their equilibrium positions should vanish. This
treatment also ensures the consistency of the static and dynamical
properties of the crystal.
\begin{table}[h!]
 \centering
\caption{Comparison of experimental and calculated IR and Raman frequencies}
 \begin{tabular}{c|c|c|c}\hline\hline\multicolumn{3}{c}{}\\ \multicolumn{4}{c}{Raman active} \\[3pt]\hline&&\multicolumn{2}{|c}{} \\
Mode&Polarization&\multicolumn{2}{c}{Frequency (cm$^{-1}$)}\\&&Experiment\cite{bi17}&Calculated\\[3pt]\hline&&&\\\multirow{8}{*}{$\rm{A_g}$}&\multirow{8}{*}{ aa}   &573 &593.30\\[3pt]
&    &497 & \_       \\
&    &460 &476.76 \\
&    &367 &375\\
&    &297 &275.9\\
&    &178 & 191.33      \\
&    &167 & \_   \\
&    &122 &108.86  \\[3pt] \hline\end{tabular} \s
 \begin{tabular}{c|c|c|c}\hline\hline\multicolumn{3}{c}{}\\ \multicolumn{4}{c}{IR active} \\[3pt]\hline&&\multicolumn{2}{|c}{} \\
Mode&Polarization&\multicolumn{2}{c}{Frequency (cm$^{-1}$)}\\&&Experiment\cite{bi14}&Calculated\\[3pt] \hline \hline&&&\\
\multirow{4}{*}{$\rm{B_{2u}}$}&\multirow{4}{*}{ b}&240   &\_  \\
 &  & 288 &  268.79 \\
& &312 &  \_  \\
& &416 &  449.88 \\[3pt] \hline&&&\\
\multirow{5}{*}{$\rm{B_{3u}}$}&\multirow{5}{*}{ a}  &240 & 222.88   \\
& & 288 & 269.621\\
& & 320 & \_ \\
& & 392 &375.54\\
& & 440 & 463.08/474.46\\[3pt] \hline\hline\end{tabular}
 \label{tb2}
 \end{table}
 It should be noted, however, that satisfying static equilibrium conditions
does not guarantee dynamical stability, namely ensuring the reality
of the phonon frequencies throughout the BZ. It should also be noted
that the shell-model parameters do not appear in the static
equilibrium equations ions are treated as rigid bodies. However,
these parameters are introduced into the dynamical matrix and are
determined through the process of fitting the 17 experimental IR and
Raman modes reported in the literature. The short-range potential
and shell-model  parameters that produce the best fit for these
modes are listed in Table I. A comparison of the experimental and
calculated values of these modes is given in Table II.

\begin{figure*}[h!]
  \includegraphics[width=6.5in]{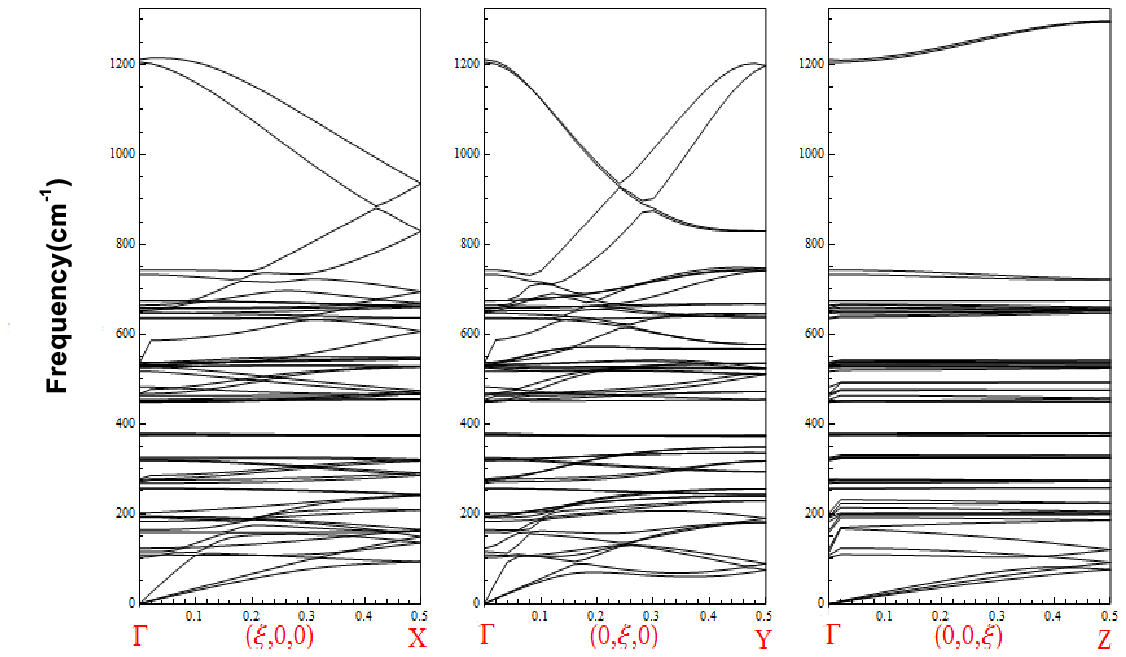}\sv \hspace{0.5in}($a$-direction)\hspace{1.in}($b$-direction)\hspace{1.3in}($c$-direction)\s
  \caption{Bulk phonon dispersion curves along {\bf a} ($\xi=q/G_a$), {\bf b} ($\xi=q/G_b$) and {\bf c} ($\xi=q/G_c$).}\label{fg3f}
\end{figure*}

The complete set of calculated bulk phonon dispersion curves (60
branches) is shown in figure \ref{fg3f}, along the high-symmetry
directions $\Gamma$-$X$, $\Gamma$-$Y$ and $\Gamma$-$Z$.

According to the point group $mmm$ ($D_{\rm 2h}$) of the space group
$Pnma$, there are 30 Raman active and 27 IR active modes at the
$\Gamma$-point, which are classified in terms of the corresponding
irreducible representations as
\begin{align*} 10\/A_{g}\oplus5B_{1g}\oplus10B_{2g}\oplus5B_{3g}\quad &\text{Raman}\\\ 9B_{1u}\oplus4B_{2u}\oplus9B_{3u}\quad&\text{IR}\end{align*}
The remaining modes at the $\Gamma$-point include five $A_{u}$ modes
and 3 zero-frequency acoustic modes. The identification of the
symmetries of the calculated modes at the $\Gamma$-point was
established with the aid of the symmetry projection
operators\cite{bi21} of the point group {\sl mmm}. Away from the
$\Gamma$-point, the point group of the wave vector is isomorphic
with $C_{\rm 2v}$ for all 3 high-symmetry directions. Symmetry
compatibility relations give\s

\hspace{0.5in}\begin{tabular}{c|c|c|c}\hline&\multicolumn{3}{c}{}\\$C_{\rm 2v}$&\multicolumn{3}{c}{$D_{\rm 2h}$}\\[3pt]\hline&&&\\& $\Delta\,(\left<100\right>)$& $\Sigma\,(\left<010\right>)$& $\Lambda\,(\left<001\right>)$\\[3pt]\hline&&&\\\multirow{2}{*}{$\text{A}_1$}&$\text{A}_g$&$\text{A}_g$&$\text{A}_g$\\&$\text{B}_{3u}$&$\text{B}_{2u}$&$\text{B}_{1u}$\\[3pt]\hline&&&\\\multirow{2}{*}{$\text{A}_2$}&$\text{A}_u$&$\text{A}_u$&$\text{A}_u$\\&$\text{B}_{3g}$&$\text{B}_{2g}$&$\text{B}_{1g}$\\[3pt]\hline&&&\\\multirow{2}{*}{$\text{B}_1$}&$\text{B}_{1g}$&$\text{B}_{1g}$&$\text{B}_{3g}$\\&$\text{B}_{2u}$&$\text{B}_{3u}$&$\text{B}_{2u}$\\[3pt]\hline&&&\\\multirow{2}{*}{$\text{B}_2$}&$\text{B}_{2g}$&$\text{B}_{3g}$&$\text{B}_{2g}$\\&$\text{B}_{1u}$&$\text{B}_{1u}$&$\text{B}_{3u}$\\[3pt]\hline\end{tabular}\s
In all, we have $\rm 20A_{1}\oplus10A_{2}\oplus10B_{1}\oplus20B_{2}$
along the $\left<100\right>$- and the $\left<001\right>$-directions,
and $\rm 15A_{1}\oplus15A_{2}\oplus15B_{1}\oplus15B_{2}$ along the
$\left<010\right>$-direction.

The highest two bands ($\simeq\,$800-1200cm$^{-1}$) are longitudinal
optic phonons involving the motion of the eight O$^{-2}$ ions in the
primitive cell; in the upper band the four O1 ions at positions
$\{0.137, 1/4, 0.405\},\,\{0.363, 3/4, 0.905\},\,\{0.863, 3/4,
0.595\},\\\{0.637,1/4, 0.095\}$ move in-phase with each other, but
anti-phase with the four O2 ions at positions $\{0.115, 1/4,
0.105\},\,\{0.385, 3/4, 0.605\},\,\{0.885, 3/4,0.895\},\\\{0.615,
1/4, 0.395\}$, while in the lower band, the O1 and O2 ions still
have anti-phase motion, but also the ions at $y=1/4$ have anti-phase
motion with those at $y=3/4$. The lowest three bands (acoustic
phonons) mix with higher bands at higher ${\bf q}$ vectors in the
$a$- and $b$-directions, but remain distinct in the
$c$-direction.\vspace{0.3in}

\begin{center}{\bf 2. Lattice Dynamics of Slab-geometry and Surface Phonons}\end{center}\s
A lattice dynamics shell-model with slab-geometry consisting of 36
layers, and terminated with  Li$^{+1}$Cu$^{+2}$O$^{-2}_2$ surfaces,
was used to calculate the phonon dispersion curves of that surface.
Since translation symmetry is broken normal to the slab surfaces,
primitive and non-primitive translations along this direction are
not allowed. Consequently, the slab symmetry reduces to {\sl 2ma},
while the surface symmetry is {\sl p1m}.

Surface equilibrium analysis showed that the Cu$^{+2}$ and Li$^{+1}$
surface ions have to be displaced outward, along the surface normal,
from its bulk position, in order to satisfy equilibrium conditions
in that direction. The pair potential and shell-model parameters of
Table I were used in the initial slab calculations. However, the
surface parameters had to be modified in order to achieve global
stability in the entire surface BZ.
\begin{figure*}[h]
    \includegraphics[width=6.5in]{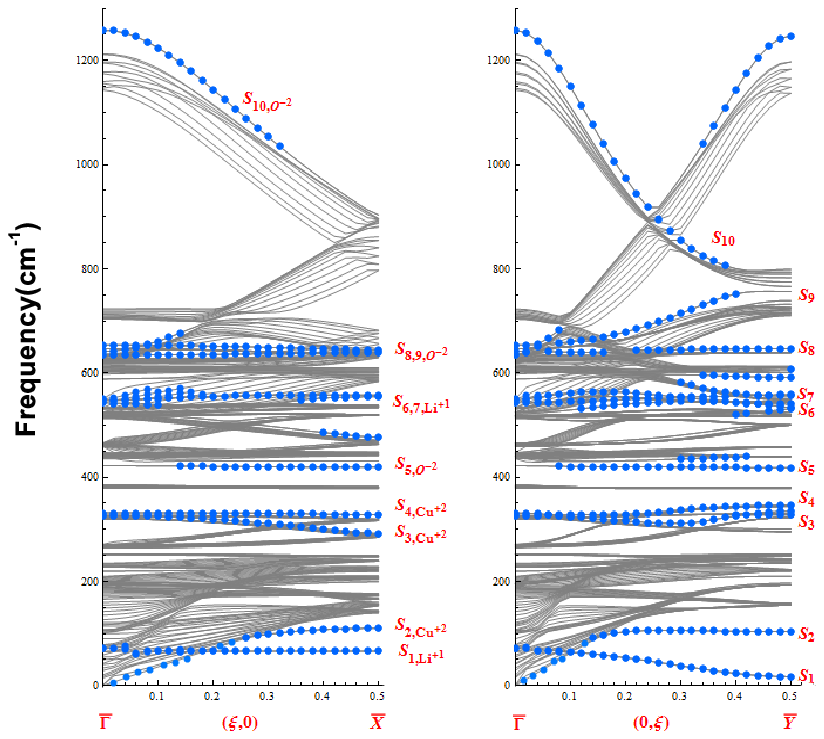}\\
  \caption{Surface phonon dispersion curves, indicated by blue dots superimposed on a gray
  background. The latter represents the surface projection of bulk
bands on the surface BZ. The surface modes are labeled
S$_{i},\,i=1,\ldots,10$.}\label{fg4}
\end{figure*}

%The lowest two bands $S_1$ and $S_2$ contain an admixture of , with polarizations in the $z$-direction. Bands $S_3$ and $S_4$ involve Cu$^{+2}$ motion. The polarizations for $S_3$ and $S_4$ along the $a$-direction are along the $b$ and $a$ directions, respectively$S_1$ and $S_2$$S_1$ and $S_2$
\begin{table}[h!]\caption{Character of the surface phonon dispersion curves. (Pol.:polarization, AP: anti-phase, IP: in-phase)}\sv
\begin{tabular}{c|c|c|c|c|c}\hline&\multicolumn{2}{c|}{}&\multicolumn{2}{c|}{}&\\\multirow{2}{*}{Branch}&\multicolumn{2}{c|}{$\left<10\right>$}&\multicolumn{2}{c|}{$\left<01\right>$}&\multirow{2}{*}{Type}\\[3pt]&Ions&Pol.&Ions&Pol.&\\[3pt]\hline&&&&&\\\multirow{2}{*}{$S_1$}&\multirow{2}{*}{ Li$^{+1}$}&\multirow{2}{*}{$z$}&Cu$^{+2}$, Li$^{+1}$&\multirow{2}{*}{$z$}&\multirow{2}{*}{AP ($\left<01\right>$)}\\[3pt]&&&1\quad:\quad5&\\[3pt]\multirow{2}{*}{$S_2$}&\multirow{2}{*}{Cu$^{+2}$}&\multirow{2}{*}{$z$}&Cu$^{+2}$, Li$^{+1}$&\multirow{2}{*}{$z$}&\multirow{2}{*}{IP ($\left<01\right>$)}\\[3pt]&&&5\quad:\quad1&\\[3pt]$S_3$&Cu$^{+2}$&$y$&Cu$^{+2}$&$x$&\\[3pt]$S_4$&Cu$^{+2}$&$x$&Cu$^{+2}$&$y$&\\[3pt]$S_5$&O$^{-2}$&$z$&O$^{-2}$&$z$&\\[3pt]$S_6$&Li$^{+1}$&$y$&Li$^{+1}$&$x$&\\[3pt]$S_7$&Li$^{+1}$&$x$&Li$^{+1}$&$y$&\\[3pt]$S_8$&O$^{-2}$&$y$&O$^{-2}$&$x$&IP\\[3pt]$S_9$&O$^{-2}$&$y$&O$^{-2}$&$y$&AP\\[3pt]$S_{10}$&O$^{-2}$&$x$&O$^{-2}$&$x$&AP\\[3pt]\hline
\end{tabular}\end{table}
 The resulting surface phonon dispersion curves are shown as blue dots superimposed on a gray background in figure \ref{fg4}; they are labeled $S_i,\ i=1,\ldots,10$. The gray background areas and lines resulting from the slab calculations correspond to contributions from the slab bulk, namely projections onto the surface BZ. Some of the dispersed gray lines would turn into solid bands as the thickness of the slab approaches infinity. The general character of the surface modes $S_1-S_{10}$ is given in Table III. $S_1$ and $S_2$ involve the motions of the Cu$^{+2}$ and Li$^{+1}$ normal to the surface. While $S_1$ is quite flat for wave vectors $q\simeq 0.15-0.5 (2\pi/a)$ along the $a$-direction, it exhibits anomalous softening in the proximity of the surface BZ boundary along the $b$-direction. This softening could be the result of the tenuous equilibrium of these ions normal to the surface.

The scattering of He beams with energies in the range 25-65 meV
allows measurement of surface phonons with frequencies bellow 400
cm$^{-1}$ (50 meV). Thus we will focus on identifying the measured
inelastic scattering events with calculated surface phonon
dispersion curves in that range. The process of identification is
further complicated by the presence of twining in the sample
crystals. A succesful procedure must be capable of sorting out modes
propagating in the $a$-direction from those propagating in the
$b$-direction. Figure 8(a) illustrates this complexity. Here, all
measured inelastic events are plotted in the left and right panels,
after reducing their momenta to the first surface BZ along the
$a$-direction (left panel) and along the $b$-direction (right
panel). Notice that each measured inelastic event has a different
value of its reduced momentum for the different directions. The
sorting criterion adopted, was to assign a given event to the
direction where it is nearest in energy to a dispersion curve. The
final result is plotted in figure 8(b). The agreement is
surprisingly quite good, especially with the two low-lying
dispersion curves. In addition, this agreement supports the surface
termination analyzed in this paper.

\begin{figure}[h!]
\hspace{-0.4in}  \includegraphics[width=4in]{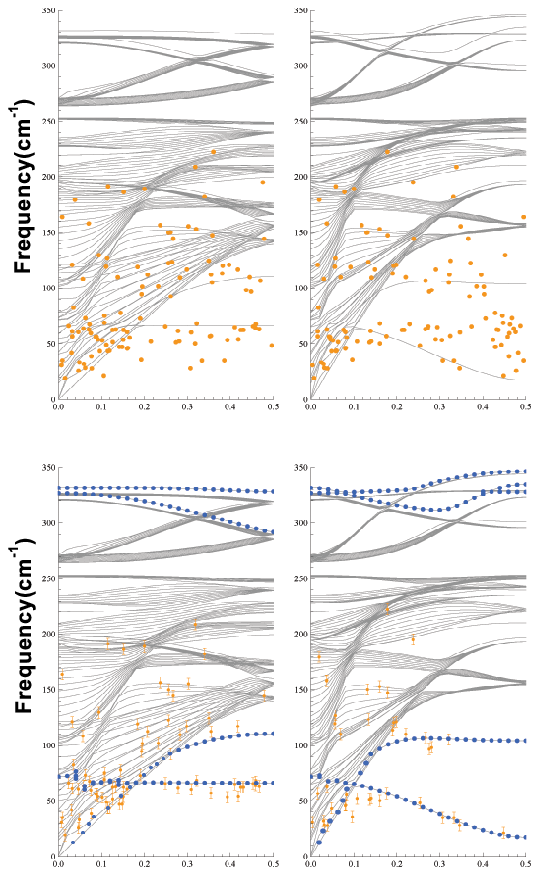}\\
  \caption{Top panels: All measured inelastic events, reduced to the proper surface BZ, superposed on calculated dispersion curves along $\left<10\right>$, (left panel), and $\left<01\right>$ (right panel). Lower panels: Fits obtained by the procedure described in the main text for $\left<10\right>$, (left panel), and $\left<01\right>$ (right panel). (Error bars are also indicated in the lower panels).}\label{fg5}
\end{figure}

In summary, with the aid of He scattering techniques at room
temperature, we find that the (001) surface of \formula is
exclusively terminated by  Li$^{+1}$Cu$^{+2}$O$^{-2}_2$, and that no
surface reconstruction occurs. Empirical fitting of lattice dynamics
shell-model based calculations to measured inelastic He scattering
spectra, supports the proposed termination and reveals that the
lowest surface phonon dispersion branches involve the motion of
Cu$^{+2}$ and Li$^{+1}$ ions normal to the surface.

\subsection*{Acknowledgement}
This work is supported by the U.S. Department of Energy under Grant
No. DE-FG02-85ER45222. FCC acknowledges the support from National
Science Council of Taiwan under project number NSC-95-2112-M-002.

\end{document}